\documentclass[aps,amsmath,amssymb,prl,showpacs,floatfix,twocolumn,superscriptaddress]{revtex4}

\usepackage{graphicx}
\usepackage{dcolumn}
\usepackage{wrapfig}
\usepackage{color}

\newcommand{\comment}[1]{}

\begin{document}

\title{Charge Fluctuations and the Valence Transition in Yb under Pressure}

\author{E. R. Ylvisaker}
\affiliation{Department of Physics, University of California, Davis, California, 95616}

\author{J. Kune\v{s}}
\affiliation{Theoretical Physics III, Center for Electronic Correlations and Magnetism,
 Institute of Physics, University of Augsburg, Augsburg 86135, Germany}
\affiliation{Institute of Physics, Academy of Sciences of the Czech Republic, Cukrovarnick\'a 10, 162 53 Praha 6, Czech Republic}

\author{A. K. McMahan}
\affiliation{Lawrence Livermore National Laboratory, Livermore, California 94550-9234, USA}

\author{W. E. Pickett}
\affiliation{Department of Physics, University of California, Davis, California, 95616}

\date{\today}

\begin{abstract}
We present a dynamical mean field theory study of the valence transition 
($f^{14} \rightarrow f^{13}$) in elemental, metallic Yb under pressure.  
Our calculations reproduce the observed valence transition as reflected 
in the volume dependence of the $4f$ occupation.  The transition is 
accelerated by heating, and suggests quasiparticle or Kondo-like structure 
in the spectra of the trivalent end state, consistent with the early 
lanthanides.  Results for the local charge fluctuations and susceptibility, 
however, show novel signatures uniquely associated with the valence transition 
itself, indicating that Yb is a fluctuating valence material in contrast to 
the intermediate valence behavior seen in the early trivalent lanthanides Ce, Pr, and Nd.
\end{abstract}

\maketitle

The valence state of rare earth atoms in lanthanide compounds has a crucial 
effect on their physical properties. Determination of the lanthanide valence
from first principles and description of the 4$f$ electrons
has been a long standing challenge due to the duality between their atomic 
character, stemming from the on-site electron-electron interaction, and the 
itinerant character, due to the lattice periodicity. 
Theories based on the two species picture, which treat part
of the 4$f$ electrons as atomic and the rest as itinerant, succeeded in reproducing the trends across the lanthanide series for compounds with
integer valence \cite{strange_nature}.  Nevertheless, besides being conceptually unsatisfactory,
the two species picture cannot describe transitions between different valence states
as well as such outstanding behavior as heavy fermions.
Only recently has the combination (LDA+DMFT) \cite{ldadmftb,ldadmftc} of the local-density approximation (LDA) and dynamical mean-field theory (DMFT) been applied to the rare earth materials \cite{Zoelfl,Held,Andy-RTS-Ce,Andy-CePrNd,Amadon,haule},
providing a material specific electronic structure including the local
many-body dynamics.

In this Letter we employ the LDA+DMFT method to study the valence transition and charge fluctuations 
in elemental Yb metal.  It is well known that Yb and Eu behave differently from the other lanthanides in their elemental 
form. If we define the valence state as the number of electrons participating in bonding,
the majority of the lanthanide series is trivalent, however for Yb and Eu the $3+$ and $2+$ valence
states are close to degenerate with $2+$ state being more stable at ambient conditions \cite{strange_nature}. 
This results in a number of anomalous properties, such as a larger molar volume (as compared to the trend the rest of
the lanthanide series follows), a lower bulk modulus \cite{Takemura85}, and the thermal expansion coefficient of Yb being three times larger than for most other lanthanides \cite{Barson57}.  The gradual transition from the divalent to the trivalent state occurs in Yb over the range 0-34 GPa, where the full $4f$ shell is opened as an $f$ electron is promoted to the valence band.  The doped $4f$ holes can move through the crystal by thermally activated hopping with $spd$ bands acting as particle and energy reservoirs, {\it fluctuating valence} behavior, or they can move coherently between atomic sites forming a narrow band well known for heavy fermions, {\it intermediate valence} behavior. We will discuss how these concepts apply to Yb.

\begin{figure}[t]
\begin{centering}
\includegraphics[width=\columnwidth]{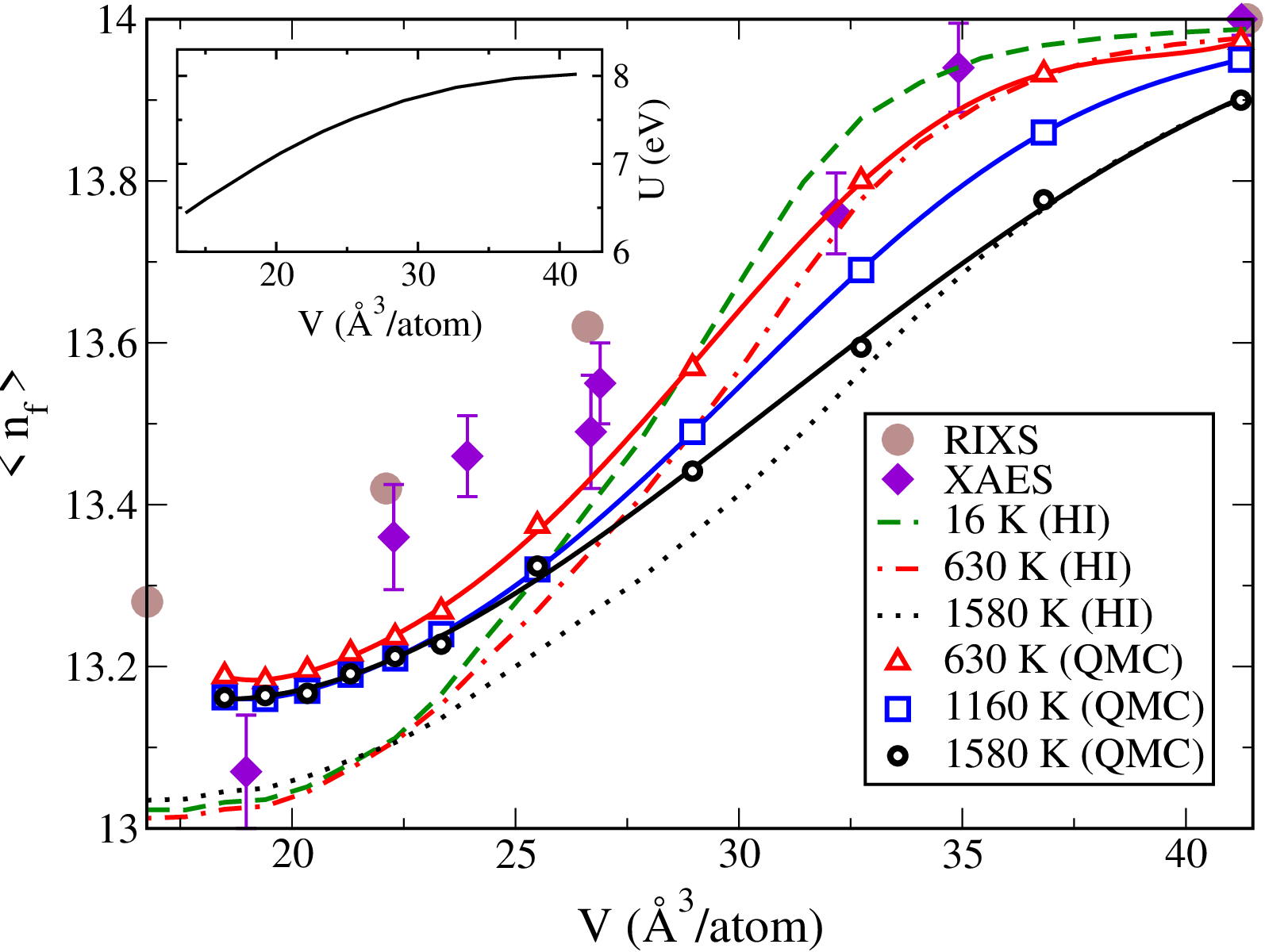}
\caption{(Color online) The $f$ shell occupation vs pressure at various temperatures compared
to the experimental data (closed symbols). Open symbols are QMC data; the 630~K and 1160~K curves were obtained with CTQMC, and the 1580~K curve with HFQMC at $L = 80$. The consistency of HFQMC with the CTQMC data was checked by $L\rightarrow\infty$ extrapolation of the $L=$80, 112 and 160 data at 630~K.  X-ray absorption edge spectroscopy (XAES)
data was taken from Ref. \onlinecite{Syassen82}, and resonant inelastic x-ray scattering (RIXS) data taken from
Ref. \onlinecite{Dallera06}.  Inset shows the interaction $U$ vs. specific volume.
}
\label{fig:nf}
\end{centering}
\end{figure}

 
In the present study we start with a self-consistent LDA calculation,
transform its one-electron Hamiltonian into an orthogonal Wannier
function basis, $H_{\bf k}^{\rm LDA}$, and then calculate the $4f$ interaction
parameter $U$ as well as the the double counting correction to $H_{\bf
k}^{\rm LDA}$.  Since the valence transition of Yb 
is of principal interest here, we have to retain the $6s$, $6p$, and $5d$ valence orbitals to allow
changes of the $f$-shell occupation. We assume the SU(N) symmetric form of the local interaction 
\begin{equation}
	H_{int}=\tfrac12 U ~{\sum_{i \neq j}} ~ \hat{n}_f^i\hat{n}_f^j,
\end{equation}
(the interaction parameter $U$ is a function of the specific volume is shown in the inset of 
Fig.\ \ref{fig:nf}), and
express the local Green's function in the relativistic $j=5/2$, $7/2$ basis to make the off-diagonal elements
small so that they can be neglected.
To solve the auxiliary impurity problem we employ two Quantum Monte Carlo (QMC) solvers, one using the Hirsch-Fye (HFQMC) 
\cite{HirschFye} algorithm and the other using the hybridization expansion continuous 
time QMC algorithm (CTQMC) \cite{Werner06} as well as an approximate, but computationally efficient Hubbard-I (HI) solver. 
Extrapolations of the HFQMC results to an infinite number of imaginary time slices $L$ were found to agree within statistical uncertainties with the CTQMC.  However, we point out that especially at the larger $L$ values, the HFQMC calculations had significant difficulties with ergodicity in the midst of the Yb valence transition where large fluctuations in $n_f$ were encountered in the Ising space sampling.  The CTQMC method did not encounter such problems, nor did HFQMC without any valence transition (e.g., Ce, Pr, and Nd \cite{Andy-CePrNd,AndyPrivate}).

In Fig.\ \ref{fig:nf} we show the evolution of the $f$-orbital occupation $n_f$ 
as a function of the specific volume $V$ at various temperatures. Comparison 
to the experimental x-ray absorption edge spectrosocpy (XAES) and resonant 
inelastic x-ray spectroscopy (RIXS) data shows some underestimation of $n_f$ at
lower volumes, possibly originating from the lack of charge self-consistency, 
which would take into account the correlation driven redistribution of charge 
between the $f$ and $spd$ orbitals.  Overall the DMFT(QMC) results agree 
reasonably well with the experimental data over the entire pressure range of 
interest.  Note that the rising $6s$ and $6p$ bands lead to an 
increase in $n_f$ with compression for the trivalent rare earths 
\cite{Andy-CePrNd}, so that it is likely that the plateau in the DMFT(QMC) 
value reached at the smallest volumes in Fig.\ \ref{fig:nf} is indeed near the 
end of the transition
from divalent to trivalent in spite of the fact that $n_f$ is
still larger than 13.0.  
The ability of DMFT(HI) to produce the valence transition seen in 
experiment suggests that the physics is essentially related to charge transfer 
between the $f$ and $spd$ orbitals. At smaller volumes where delocalization of 
$f$-electrons becomes more important HI overestimates the number of holes in 
the $4f$ shells. 

The calculated $n_f$ values show a sizable temperature dependence both with HI
 and QMC, particularly at low pressure 
as is evident in Fig.\ \ref{fig:nf}.  The general trend is that at higher temperatures, 
the $f^{13}$ state is favored over the $f^{14}$ state.  This trend can be 
followed down to the lowest studied temperature of 16~K, 
which is only accessible with HI.  We expect this trend would hold at lower temperatures for QMC calculations as well.  Temperature dependent 
measurements of the valence of Yb in YbInCu$_4$ show the same trend 
($4f$ occupation decreasing with increasing temperature) \cite{Reinert98}, 
with different studies
finding a minimum of about 13.1 \cite{Sato04,Schmidt05}. The same behavior 
was found for YbAgCu$_4$ and YbAl$_3$ \cite{Joyce96}.  
The authors of Ref. \onlinecite{Joyce96} did not find any significant 
difference in elemental Yb between
the temperatures of 250~K and 25~K; at 250~K Yb should be already strongly 
divalent and 
lowering the temperature cannot decrease the valence any further.  There is, 
however, a significant broadening 
of the spectrum at higher temperature, so we expect that the trend seen in 
our calculations would be borne out 
if measurements were carried out at higher temperatures.  At high pressure 
the temperature dependence of $n_f$ becomes weak.  
The chemical environment of Yb can affect the temperature sensitivity of 
$n_f$ as well; in fact, 
the measured valence of Yb in YbGaGe has been recently found to be 
temperature independent \cite{Doyle07}, 
and Yb nearly divalent. This is somewhat a contrast to our result here, 
where the temperature sensitivity is only 
near the divalent state. The details of the electronic structure of 
YbGaGe that causes 
the Yb valence to be temperature independent are not understood. 
\begin{figure}[t]
\begin{centering}
\includegraphics[width=0.45\textwidth]{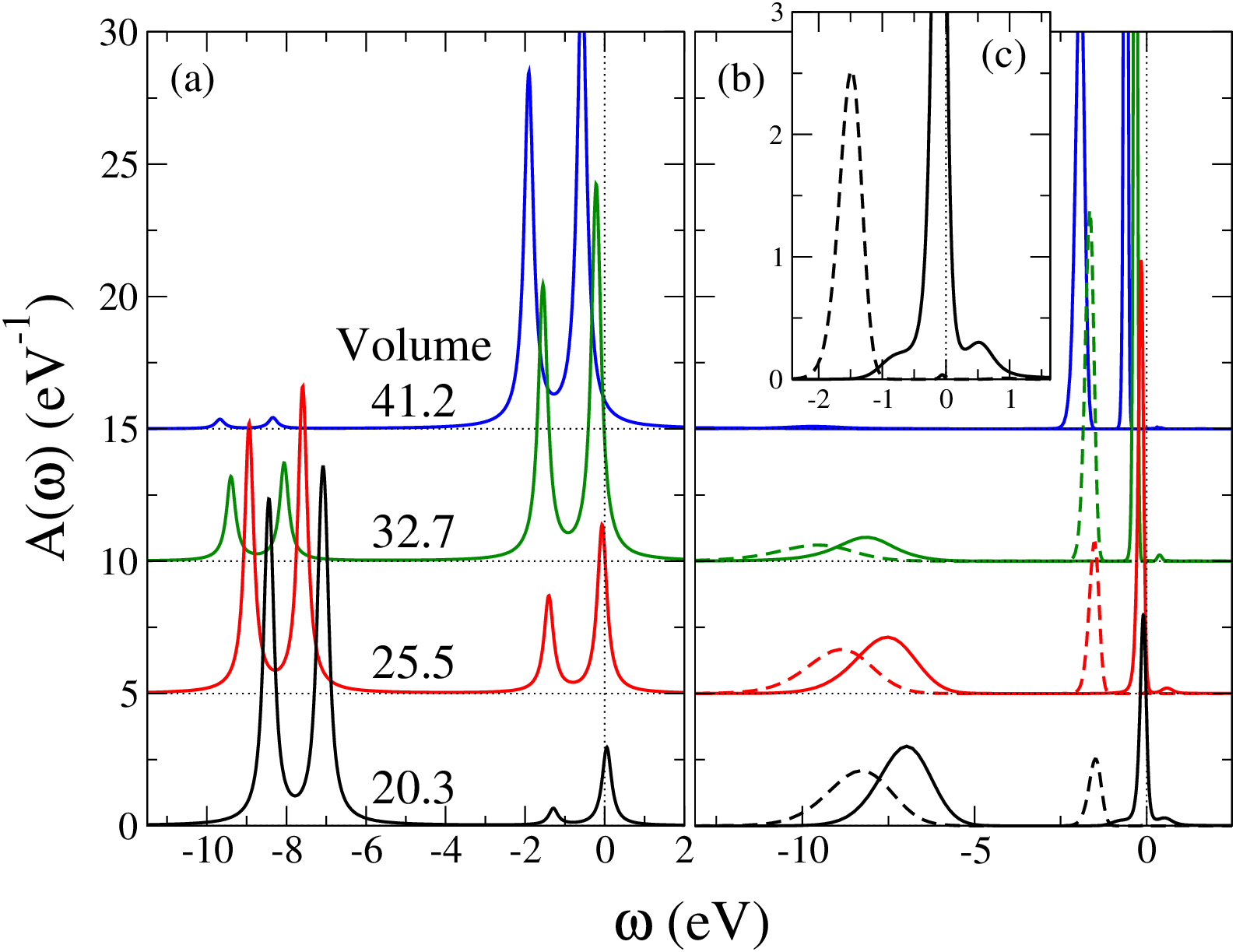}
\caption{(Color online) Plot of Yb spectral function $A(\omega)$ calculated 
in (a) HI at $T = 630$ K and
(b) CTQMC at $T  = 1160$ K for selected volumes (in \AA$^3$/atom).
Dashed curves in (b) are for $j = \tfrac52$; solid curves, $j = \tfrac72$.
Spin-orbit split peaks corresponding to the $f^{14} \rightarrow f^{13}$
excitation are near the chemical potential, while those corresponding
to the $f^{13} \rightarrow f^{12}$ excitation are around $-U(V)$.
An expanded view near the Fermi-level for V = 20.3 \AA$^3$/atom is shown in (c).
The $j=\tfrac72$ curve shows two shoulders around the large central peak.  }
\label{fig:Hub1:spct}
\end{centering}
\end{figure}

The calculated $4f$ spectral densities $A(\omega)$ are shown in Fig.\ \ref{fig:Hub1:spct}, as obtained by 
direct evaluation of the Green's function near the real frequency axis for 
HI, and applying the 
maximum entropy method \cite{SilverSiviaGubernatis} to the CTQMC data. At 
low pressure, there is a clear spin-orbit-split peak near the chemical 
potential ($\omega=0$),
which corresponds to the $f^{14} \rightarrow f^{13}$ excitation. As the 
pressure is increased, the weight of this double-peak decreases, 
and the $f^{13} \rightarrow f^{12}$ excitation appears around 8 eV 
($\approx U$) below the $f^{14} \rightarrow f^{13}$ double-peak.
While the analytic continuation smears the high energy features of the 
CTQMC spectra, HI is known to incorrectly reduce
the width of the Hubbard bands. 

%
\begin{figure}[t]
\begin{centering}
\includegraphics[width=\columnwidth]{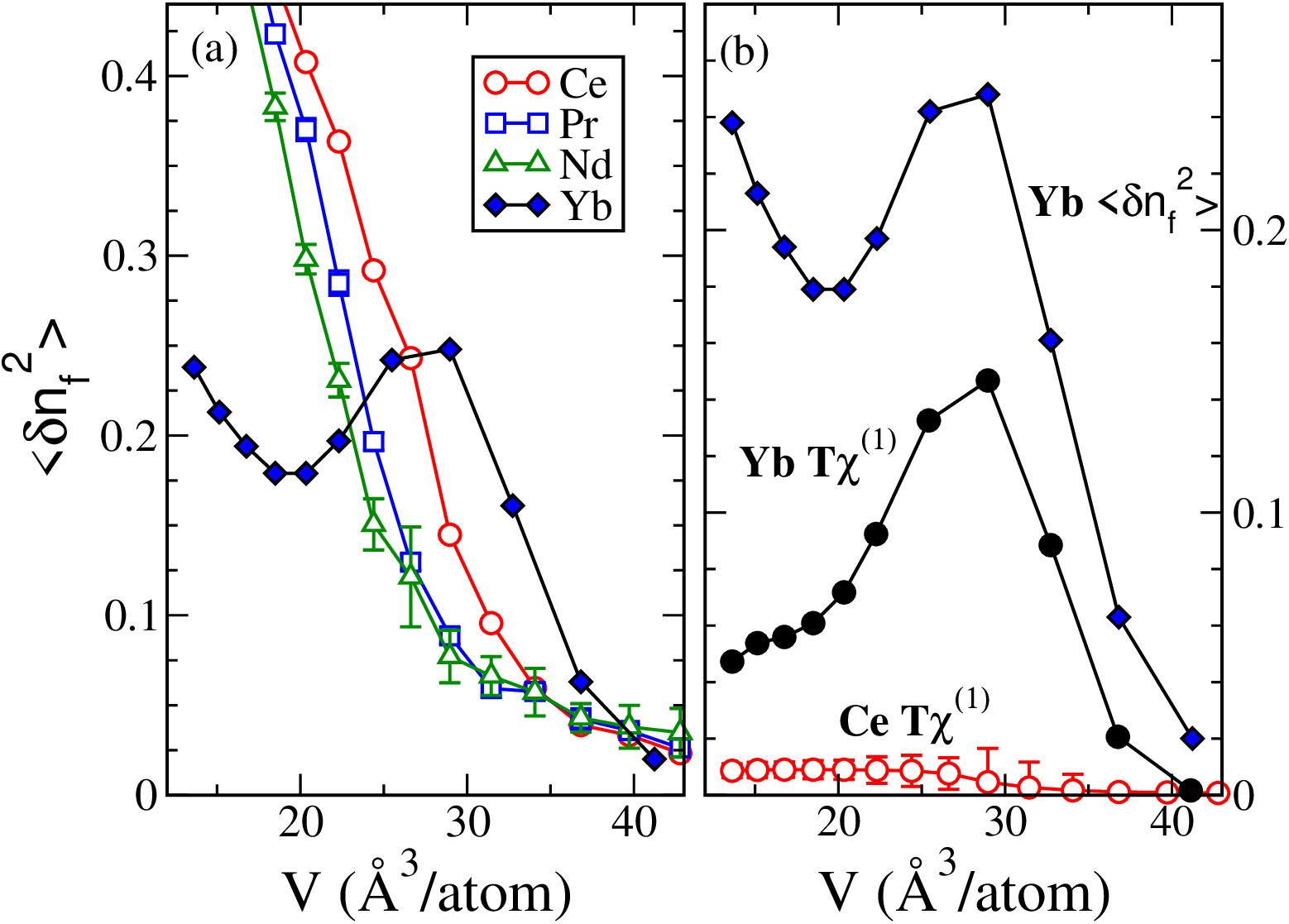}
\caption{(Color online) (a) Equal time charge fluctuations in Yb (CTQMC), Ce, Pr and Nd (HFQMC with $L \rightarrow \infty$ extrapolation) at 630 K.  
If not shown error bars are smaller than the symbol size.   
(b) Charge fluctuations for Yb (diamonds, the same as in left panel) compared
to their imaginary time average, $T\chi^{(1)}$, for Yb (closed circles) and Ce (open circles). 
 }
\label{fig:fluctuations}
\end{centering}
\end{figure}

When the valence transition approaches completion as in the HI result (bottom of Fig.\ \ref{fig:Hub1:spct}a), 
the $j = 7/2$ $f^{14} \rightarrow f^{13}$ peak overlapping the Fermi level shifts increasingly above this 
energy to become the unoccupied upper Hubbard band for the $4f$ states, corresponding to the new $4f$ hole.  
A close examination of the CTQMC counterpart (Fig.\ \ref{fig:Hub1:spct}c), on the other hand, 
shows the corresponding $j = 7/2$ structure 
to contain a relatively large peak at the Fermi level with shoulders to
either side.  Based on the temperature scaling of the Fermi-level peak,
we believe it to be a quasiparticle or Kondo-like contribution consistent
with increasing delocalization of the $4f$ hole as volume is reduced,
and that the lower shoulders are likely of more Hubbard character.
Moreover, any transfer of spectral weight into such a quasiparticle contribution 
at the expense of the $f^{13} \rightarrow f^{12}$ Hubbard band would also contribute to 
a larger DMFT(QMC) $n_f$ than in DMFT(HI) as is observed.  Finally, both the simultaneous existence of two lower 
Hubbard bands, and the gradual transfer of spectral weight evident in Fig.\ \ref{fig:Hub1:spct} are fundamental 
signatures of electron correlation which a single-particle approach such as LDA or LDA+U cannot reproduce.  

Thus far we have shown a gradual emptying of the 4$f$ shell with applied pressure and the corresponding
changes in the single-particle spectra interpreted in terms of the relative abundance of $f^{13}$ and $f^{14}$
configurations in the system. We now present an analysis of the associated local
charge fluctuations in Yb. In Fig.\ \ref{fig:fluctuations}a we compare the local charge fluctuations in Yb,
defined through the mean square deviation of the $f$-occupation $\langle \delta \hat{n}_f^2 \rangle$;
$\delta \hat{n}_f=\hat{n}_f-n_f$, with those in Ce, Pr and Nd \cite{AndyPrivate}. While Ce, Pr and Nd
show a monotonic increase of $\langle \delta \hat{n}_f^2 \rangle$ with pressure, 
the charge fluctuations in Yb exhibit a local maximum around
27~\AA$^3$/atom where $n_f \approx 13.5$.  Evaluating the thermal expectation in a basis of $\hat{n}_f$ eigenstates yields
$\langle \delta \hat{n}_f^2 \rangle= w_{13}(1\!-\!w_{13})$, if we assume only
$f^{13}$ and $f^{14}$ configurations have significant weight
($w_{13}\!+\!w_{14}=1$), and thus a peak in the midst of the valence transition
$w_{13}=0\rightarrow 1$.  On the other hand, if a single $f^n$ configuration predominates
($w_n\sim 1$) but then loses weight to $f^{n\!-\!1}$ and $f^{n\!+\!1}$ states due
to growing hybridization under compression, one will see a monotonic increase in
$\langle \delta \hat{n}_f^2 \rangle$.  It would appear likely that the behavior
seen in Fig.\ \ref{fig:fluctuations}a for Yb is reflects a combination of both
effects, while that of Ce, Pr, and Nd is primarily the latter delocalization
behavior.

\begin{figure}[t]
\begin{centering}
\includegraphics[width=\columnwidth]{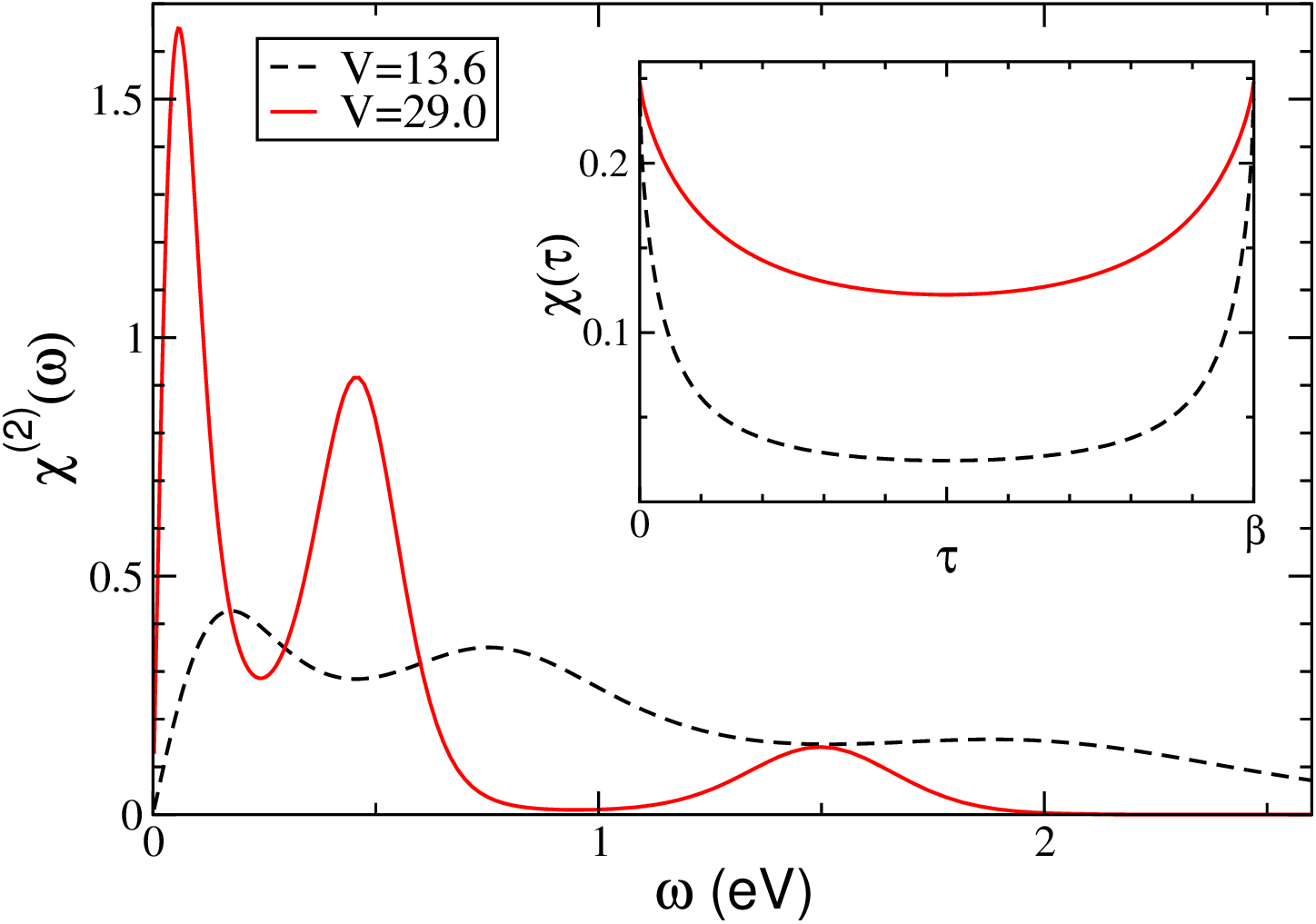}
\caption{\label{fig:chi_tau}(Color online) Imaginary part of the local $ff$ 
charge susceptibility at 630~K and specific volumes of 13.6~\AA$^3$ (black 
dashed line) and 29.0~\AA$^3$ (red solid line) per atom.  The inset shows 
the corresponding imaginary time correlation functions obtained with CTQMC.
}
\end{centering}
\end{figure}

To gain more insight we further evaluate the local charge susceptibility $\chi^{(1)}$ 
(Fig.\ \ref{fig:fluctuations}b),
obtained from the imaginary-time correlation function
\begin{equation}
	\chi^{(1)}=\int_0^{\beta}d\tau~\chi(\tau),
		\quad\chi(\tau)=\langle\delta\hat{n}_f(\tau)\delta\hat{n}_f(0)\rangle.
\end{equation}
We find that the initial increase of the charge fluctuations with pressure
is mirrored by an increase of the local charge susceptibility with both having
their maxima at about the same volume, however, the increase of 
$\langle \delta \hat{n}_f^2 \rangle$ at high pressure is not reflected in the
charge susceptibility. The origin of this behavior is revealed in the inset of
Fig.\ \ref{fig:chi_tau} where we compare the imaginary time charge correlations 
for $V$=13.6 and 29.0~\AA$^3$/atom.  While the magnitude of the charge 
fluctuations is about the same ($\tau=0$ intercepts), the fluctuations at high 
pressure are short lived (rapid decay with increasing $\tau$) leading  to a 
relatively lower susceptibility. In Fig.\ \ref{fig:chi_tau} we show the 
corresponding physical susceptibilities on the real frequency axis, imaginary 
parts of which characterizes the density of charge excitations.  
\begin{equation}
\chi(\tau)=\frac{1}{\pi}\int_0^{\infty}d\omega~\frac{e^{-\tau\omega}+e^{-(\beta-\tau)\omega}}{1-e^{-\beta\omega}}~\chi^{(2)}(\omega)
\end{equation}
Comparison of $\chi(\tau)$ and $\chi^{(2)}(\omega)$ shows that slowly decaying fluctuations are, as expected,
related to low energy charge excitations.

Low energy local charge excitations in strongly correlated metals are rather 
rare since they are usually suppressed by the onsite Coulomb repulsion, as is 
the case for Ce, Pr, and Nd where $T\chi^{(1)}$ is 20-60 times smaller than 
$\langle \delta \hat{n}_f^2 \rangle$ over the volume 
range of Fig.\ \ref{fig:chi_tau}b \cite{AndyPrivate}.  Existence of such 
excitations and the corresponding maxima in $\chi^{(1)}$ vs $V$ and 
$\langle \delta \hat{n}_f^2 \rangle$ vs $V$ curves distinguishes this 
fluctuating valence behavior of Yb from the intermediate valence behavior of 
the other studied lanthanides. The calculated local charge susceptibility 
provides a clear physical meaning to the earlier reports of near degeneracy 
of $2+$ and $3+$ valence states in Yb \cite{strange_nature}. There is also a 
direct connection to the experimental observation of the unusual 
pressure-volume curve of Yb, namely the compressibility of the electronic 
subsystem is proportional to its charge susceptibility.  It is plausible to 
expect that the maximum of local $\chi^{(1)}$ vs $V$ leads to softening of 
the entire electronic liquid, although the effect will be to some extent 
masked by the presence of $spd$ bands.

Our local charge fluctuation and susceptibility results suggest that pressure
affects metallic Yb essentially in two ways which can, to a first approximation,
be looked at separately: {\it hole doping} of the filled $4f$ shell and {\it
growing hybridization} of the $4f$ bands with the $spd$ bands.  In regard to 
the first mechanism, it
is the crossover of atomic-like $f^{13}$ and $f^{14}$ levels which drives the
valence transition creating a peak in $\langle \delta \hat{n}_f^2 \rangle$ as
noted.  Since the system is then also sensitive to external perturbations breaking
this near degeneracy, a large charge susceptibility also follows.  Hole doping
into the atomic-like $4f$ shell can therefore explain both the (local) maxima in
$\langle \delta \hat{n}_f^2 \rangle$ and $\chi^{(1)}$.  The second mechanism,
growing hybridization with its concomitant $4f$ delocalization, leads to an
increase of charge fluctuation $\langle \delta \hat{n}_f^2 \rangle$ at higher
pressures, as well as to the screening of these fluctuations, and thus the small
charge susceptibility $\chi^{(1)}$, in analogy to screening of magnetic moment in
the Anderson impurity model.

We have examined the pressure induced valence transition of elemental
Yb using the LDA+DMFT approach.  The transition is advanced also by 
increasing temperature at larger volumes, and appears to have
reached its trivalent limit by the smallest volumes considered,
where there is evidence of Kondo-like structure in the spectra as
found with the early lanthanides.  Perhaps most interestingly we
find that while the $f$ orbital occupation and the single-particle 
spectra evolve monotonically with compression, the charge fluctuations
and local charge susceptibility
exhibit a distinct peak. We interpret this feature as fluctuating 
valence behavior in contrast to Ce, Pr and Nd 
which can be classified as intermediate valence systems.

The authors would like to thank Philipp Werner for providing his CTQMC code and Richard T. Scalettar and Simone Chiesa for stimulating discussions.  
E. R. Y. and W. E. P were supported by DOE SciDAC Grant No. DE-FC02-06ER25794.  J. K. acknowledges the support of SFB 484 of the Deutsche Forschungsgemeinschaft.  
Work at LLNL was performed under the auspices of the U.S. DOE under contract W-7405-Eng-48.

\bibliographystyle{apsrev}
\bibliography{Yb}

\end{document}